# Analytical Model for the Optical Functions of Indium Gallium Nitride with Application to Thin Film Solar Photovoltaic Cells

Dirk V. P. McLaughlin[1] and Joshua M. Pearce[2]*
1. Department of Mechanical and Materials Engineering, Queen's University, Canada
2. Department of Materials Science & Engineering and Department of Electrical & Computer Engineering, Michigan Technological University, U.S.A.

**Contact author:**

Joshua M. Pearce
601 M&M Building
Michigan Technological University
1400 Townsend Drive
Houghton, MI 49931-1295
906-487-1466
pearce@mtu.edu

## Abstract

This paper presents the preliminary results of optical characterization using spectroscopic ellipsometry of wurtzite indium gallium nitride ($In_xGa_{1-x}N$) thin films with medium indium content ($0.38<x<0.68$) that were deposited on silicon dioxide using plasma-enhanced evaporation. A Kramers-Kronig consistent parametric analytical model using Gaussian oscillators to describe the absorption spectra has been developed to extract the real and imaginary components of the dielectric function ($\varepsilon_1$, $\varepsilon_2$) of $In_xGa_{1-x}N$ films. Scanning electron microscope (SEM) images are presented to examine film microstructure and verify film thicknesses determined from ellipsometry modelling. This fitting procedure, model, and parameters can be employed in the future to extract physical parameters from ellipsometric data from other $In_xGa_{1-x}N$ films.

# 1. Introduction

While indium gallium nitride (In$_x$Ga$_{1-x}$N) semiconductors have been used in blue and ultraviolet LEDs for many years [1,2], only recently has significant research been performed on the material for solar photovoltaic (PV) applications. The optoelectronic properties of In$_x$Ga$_{1-x}$N make the material very well-suited for use in PV because the band gap of In$_x$Ga$_{1-x}$N can be tuned from 0.7 eV to 3.4 eV by altering the indium content (*x*) in the material [3-7]. However, indium nitride (InN) and gallium nitride (GaN) have a lattice mismatch of 10% which results in phase segregation and poor quality In-rich films under most growth conditions with conventional MBE or MOCVD methods [3]. As a result, a number of optical characterization studies have been performed on In$_x$Ga$_{1-x}$N alloys with low indium contents (x) [7-11] but few on alloys with x>0.2 [12]. Fortunately, these alloy films can be created using a new plasma-enhanced evaporation deposition process, which is well-suited for large-scale manufacturing of photovoltaic devices because it can deposit In$_x$Ga$_{1-x}$N on silicon dioxide [13]. This paper presents the results of optical characterization using spectroscopic ellipsometry of wurtzite In$_x$Ga$_{1-x}$N thin films with medium indium content (0.38<x<0.68) and proposes an analytical model for the optical functions of In$_x$Ga$_{1-x}$N.

Optical characterization via ellipsometry to determine the optoelectronic properties of a material for device design is extremely valuable given a reliable analytical model. In this paper, a Kramers-Kronig consistent parametric model using Gaussian oscillators to describe the absorption has been developed to obtain the real and imaginary components of the dielectric function ($\varepsilon_1$, $\varepsilon_2$) of In$_x$Ga$_{1-x}$N films. Gaussian oscillators, while typically used for amorphous materials, can be used in the modelling of a wide range of materials that are transparent in some portion of the spectral range. This owes to the fact that $\varepsilon_2$ rapidly approaches zero on either side of the absorption peak. This fitting procedure and model can be employed on In-rich In$_x$Ga$_{1-x}$N alloys for optimization in solar photovoltaic cells.

# 2. Experimental

## 2.1 Sample Preparation

The In$_x$Ga$_{1-x}$N films were deposited using a plasma-enhanced evaporation process, which has been previously described in detail [13]. First, 6 nm amorphous GaN layers were deposited on 2 μm silicon dioxide coatings on silicon wafers. On top of this layer, In$_x$Ga$_{1-x}$N films with an approximate thickness of 200 nm were grown at a rate of 1.3 Å/s with a substrate temperature between 470°C and 500°C for the films analyzed in this paper.

The ratio, x, of indium to gallium was altered by changing the temperature of the metal sources. This film composition was determined by the Tauc method using absorption data obtained from the spectroscopic ellipsometry models. In this method for interband transition [14], a Tauc plot is created relating the absorption coefficients ($\alpha$) with photon energy (E) by plotting $(\alpha E)^{1/2}$ versus E. The optical band gap is equal to the x-intercept found by extrapolating the linear portion of the curve to the energy-axis. Once the optical band gap is found, the indium content can be closely approximated using the Bowing equation [3,15,16] shown in Equation 1:

$$E_g(In_xGa_{1-x}N) = E_{g,InN}x + E_{g,GaN}(1-x) - bx(1-x) \quad [eV] \quad (1)$$

where, $E_g(In_xGa_{1-x}N)$ refers to the band gap of $In_xGa_{1-x}N$ in eV, $E_{g,In}$ is the band gap of InN (0.7 eV) and $E_{g,Ga}$ is the band gap of GaN (3.4 eV). A bowing parameter, b, of 1.43 eV was found to best approximate wurtzite $In_xGa_{1-x}N$ over the entire compositional range [16,17].

## 2.2 Ellipsometric Measurements

The $In_xGa_{1-x}N$ films were characterized via spectroscopic ellipsometry using a J.A. Woollam Co. vertical-variable angle spectroscopic ellipsometer (V-VASE). An AutoRetarder was used that improves the data measurement accuracy. Two angles of incidence were chosen: 70° and 78° (centred on the 74° Brewster angle of the silicon substrate). At each angle, measurements were taken at 0.025 eV intervals within a photon energy range of 0.8 eV to 4.50 eV.

## 2.3 Parametric Modelling

Ellipsometric measurements of the $In_xGa_{1-x}N$ films yield two parameters for each wavelength and angle of incidence: Ψ and Δ, representing the change in amplitude ratio and change in phase shift, respectively, between the p- and s- components of the light beam's electric field. These raw data parameter curves are generally not useful themselves, but must be modelled through a regression-based data analysis in order to extract physical information about the thin films. Through this method important information such as film thickness, surface roughness, optical constants (dielectric constants: $\varepsilon_1$ and $\varepsilon_2$; index of refraction: η and extinction coefficient: κ) and absorption coefficients, α, can be obtained [18]. For this modelling procedure, J.A. Woollam Co.'s WVASE32 software was used.

In developing the parametric model, each unique layer of material in the sample must be represented in the model and was built on a J.A. Woollam Co. standard tabulated Si and associated $SiO_2$ substrate. The next layer on the sample (bottom up) is the 6 nm GaN buffer layer which was found to have no effect on the model fit (zero thickness) for any of the samples run so it was excluded as an individual layer. Its true thickness would be automatically included in the subsequent $In_xGa_{1-x}N$ layer. While the $SiO_2$ substrate and GaN buffer layer affect the growth of the $In_xGa_{1-x}N$ layer, the light interaction and photon absorption is dominated by the highly absorbing $In_xGa_{1-x}N$ film as revealed through the modelling process.

Since the microstructure and composition of the $In_xGa_{1-x}N$ films are unique to the deposition process and growth conditions, their optical constants cannot be represented by standard tabulated values and parametric dispersion relationships must be used. For this reason, the ellipsometric output, Ψ and Δ, were restricted to the non-absorbing regions (this would vary depending on the band gap of the $In_xGa_{1-x}N$ films) so that a Cauchy layer with Urbach absorption (imperfect crystal with some absorbing states below its band gap) could be used for this portion and a first thickness estimate is obtained. Once a strong fit was obtained, the thicknesses and Cauchy parameters were fixed and a point-by-point fit of the Cauchy layer optical constants was performed starting from the longest wavelengths. In order to produce a real physical shape for dispersion over the entire range (absorbing regions now included), the Cauchy layer was converted into a general oscillator layer to enforce Kramers-Kronig consistency. Since the absorption did not fully reach zero below the band gap, two or three Gaussian oscillators were found to fit the data very well. For most samples, the model fit was improved by allowing the optical properties (such as index of refraction, n) of the $In_xGa_{1-x}N$ film oscillator layer to vary with depth. Using this gradient method, the $In_xGa_{1-x}N$ layer is divided into multiple sublayers so that changes can be accounted for in film composition or microstructure that alter the material's optical properties. The gradient trend shows a slight increase in the index of refraction from substrate to surface as film indium content increases. This is attributed to fewer void spaces at the surface (more tightly packed 3-D islands/nanocolumns) at the surface – a phenomenon which has been previously reported [13].

To account for the optical influence of surface roughness, a fourth layer was added to the model. This layer is treated as a Bruggeman Effective Medium Approximation (EMA) consisting of a 50/50 mixture of $In_xGa_{1-x}N$ film and void space (from an optical standpoint) [19]. The fit quality was further improved by including a thickness non-uniformity option (under 2%) to account for variations in light reflection and capture due to an uneven film surface.

### 2.4 Scanning Electron Microscopy

The improvement in fit quality from the surface roughness model layer indicated an uneven surface layer which was confirmed through scanning electron microscopy (SEM) imaging. A LEO 1530/Zeiss Field Emission microscope was used at an acceleration voltage of 3.5 kV. Additionally, cross-sectional imaging of the $In_xGa_{1-x}N$ films was performed to confirm the thicknesses determined by the ellipsometry model.

## 3. Results and Discussion

### 3.1 SEM Imaging

Figure 1 presents top-down SEM images for four $In_xGa_{1-x}N$ films grown by plasma-enhanced evaporation with estimated indium contents ranging from x=0.38 to x=0.68.

{Insert Fig 1}

As Figure 1 illustrates, the surface microstructure was influenced by the indium content of the films. In Figure 1a, the 177 nm-thick $In_{0.38}Ga_{0.62}N$ film exhibits 33 nm diameter "coffee bean-like" grains with a single phase. A similar surface is seen in Figure 1b although the 202 nm-thick $In_{0.54}Ga_{0.46}N$ film has slightly larger 40 nm grains. Please note that in order to more clearly show the microstructure, Figure 1b-d have the same magnification which is slightly less than Figure 1a. However, SEM images revealed a different surface microstructure for the two films with the highest indium contents. Figures 1c and 1d for the 204 nm-thick $In_{0.64}Ga_{0.36}N$ and 221 nm-thick $In_{0.68}Ga_{0.32}N$ films show single-phase isolated platelet/nanocolumnar grains with average diameters of 85 nm and 105 nm, respectively. The $In_xGa_{1-x}N$ films display a clear relationship of larger grain or nanocolumn sizes with increasing indium content, which has been reported previously [13].

Figure 2 presents a cross-sectional SEM image for the $In_{0.68}Ga_{0.32}N$ film (top-down image in Figure 1d) revealing a partially coalesced nanocolumnar microstructure. The image also verifies the accuracy of the film thickness (221 nm) determined from the ellipsometric model.

{Insert Fig 2}

### 3.2 Extracting the Optical Constants

The Kramers-Kronig consistent parametric dispersion model described earlier was found to fit the experimental data ($\Psi$ and $\Delta$) very well with mean-squared errors (MSE) under 16. This, in addition to film thicknesses confirmed by SEM imaging, allowed the conclusion that the model developed is both accurate and physically meaningful. Using these models for the $In_xGa_{1-x}N$ films, film thicknesses and optical properties were obtained. Table 1 summarizes the estimated indium contents, $In_xGa_{1-x}N$ film thicknesses, and Bruggeman layer thicknesses determined from the ellipsometry models for each sample.

**{Insert Table 1}**

Figure 3 shows the real and imaginary parts of the dielectric function, $\varepsilon_1$ and $\varepsilon_2$, as a function of photon energy for the analyzed $In_xGa_{1-x}N$ films:

**{Insert Fig 3}**

The $In_xGa_{1-x}N$ films show similar trends for the real and imaginary parts of the dielectric function. As expected, the $\varepsilon_2$ curves indicate strong absorption at the higher photon energies above the band gaps. However, below the band gaps $\varepsilon_2$ does not reach zero as it should for a perfectly mono-crystalline semiconductor. Instead, the $\varepsilon_2$ curve increases at the low end of the photon energy range for each sample. While some sub-gap absorption a high density of absorbing free electrons, dislocations and defects in the $In_xGa_{1-x}N$ films is expected, this increase in the real part of the dielectric function could also be a deviation of the ellipsometry models at such low energies. This phenomenon can also be seen in the absorption coefficient curves shown in Figure 4. The optical constant ($\varepsilon_2$, n) curves also lack the sharp features/slopes seen in $In_xGa_{1-x}N$ films of similar thicknesses deposited by other methods [8,10], indicating that the films presented here are not perfectly crystalline. However, those $In_xGa_{1-x}N$ films were grown with low indium contents (x<0.2). Therefore, the increased broadening of the dielectric functions and index of refraction curves can be largely attributed to the greater lattice mismatch the In-rich $In_xGa_{1-x}N$ films presented here suffer from which tend to cause lattice disorder and slight fluctuations in composition. One noticeable difference between the samples is seen in the two most In-rich films in Figures 3c and 3d as these exhibit additional absorption peaks around 2.5-2.8 eV. Further examination is required to determine the exact source of this absorption.

Figure 4 shows the absorption coefficients as a function of photon energy for the $In_xGa_{1-x}N$ films:

**{Insert Fig 4}**

Large absorption coefficients are another attribute of the direct band gap $In_xGa_{1-x}N$ alloys. Figure 4 shows absorption coefficients as a function of photon energy for each sample, which are found to be on the order of $10^5$ cm$^{-1}$ at the absorption edges. This strong absorption, despite imperfect crystallinity, illustrates the potential of the $In_xGa_{1-x}N$ for photovoltaic applications.

Typically, transmission and reflection (T&R) measurements are performed to obtain absorption information [20]. However, spectroscopic ellipsometry is a good alternative if T&R is not possible due to opaque substrates such as the $Si/SiO_2$ used here. Additionally, ellipsometry can be used to determine the optical band gap of a thin film if photoluminescence and spectrophotometry are not options.

### 3.3 Ellipsometry Model Parameters

The parameters in the analytical expression for the optical functions of $In_xGa_{1-x}N$ films are presented for the benefit of other researchers of $In_xGa_{1-x}N$ films (by sputtering or other methods) to fit ellipsometric data for the extraction of physical information. Second, the expressions can be used in performance simulations of solar photovoltaic cells, in particular the multi-junction devices $In_xGa_{1-x}N$ has been theorized to be so well-suited towards.

In this paper, the same Gaussian oscillator-based model is used to fit the experimental data for each $In_xGa_{1-x}N$ film. However, the parameters of the oscillators will differ for each depending on the film's thicknesses and optical properties. Table 2 presents parameters for the Gaussian oscillators used in the modelling of each film. The Gaussian oscillator equation as it relates to $\varepsilon_2$ is shown below as Equation 2:

$$\varepsilon_2(E) = A \cdot [\exp(-((E-E_n)/\sigma)^2) - \exp(-((E+E_n)/\sigma)^2)] \qquad (2)$$

*where A* refers to the amplitude of the curve's peak, $E_n$ refers to the centring energy or energy at the curve's peak, and $\sigma$ is the standard deviation of the curve.

**{Insert Table 2}**

As a result of sub-gap absorption, each $In_xGa_{1-x}N$ film requires one oscillator centred at very low energy levels in addition to an oscillator describing absorption of photon energies above the band gap. The extra absorption peaks of the most indium-rich $In_{0.64}Ga_{0.36}N$ and $In_{0.68}Ga_{0.32}N$ films are modelled using an additional small oscillator centred at about 2.6 eV. The size, location and number of oscillators will vary for every thin film depending on the composition and complexity of the film's microstructure. However, the parameters presented can be used to help describe the optical functions and absorption of similar wurtzite In-rich $In_xGa_{1-x}N$ films. This is particularly important for the complex nature of proposed multi-junction solar photovoltaic devices, which rely on multiple layers of $In_xGa_{1-x}N$ films with varying indium content [6,21,22].

## 4. Conclusions

A Kramers-Kronig consistent parametric model has been developed for the optical functions of wurtzite $In_xGa_{1-x}N$ alloy films of medium indium contents (0.38<x<0.68) deposited by a novel plasma-enhanced evaporation deposition system. This model employing simple Gaussian oscillators was used to fit spectroscopic ellipsometric data over the 0.8 eV to 4.5 eV range to obtain film thicknesses, dielectric functions and absorption coefficients. Using analytical expressions to accurately describe the optical functions of $In_xGa_{1-x}N$ films is an extremely important step in understanding the semiconductor and its utilization in high-efficiency solar photovoltaic cells. The optical characterization methods employed and the model developed can be used as a basis for the optical characterization of similar $In_xGa_{1-x}N$ thin films.

## Acknowledgements

This work was supported by the Natural Sciences and Engineering Research Council of Canada, Canada Foundation for Innovation, Ministry of Research and Innovation, and J.A. Woollam Co. The authors would also like to acknowledge helpful discussions with M. Einav, C. Elliot, and R. Synowicki.

**Vitae**

Dirk V. P. McLaughlin is a member of the Queen's Applied Sustainability Research Group in Ontario, Canada. His research concentrates on the use of indium gallium nitride for solar photovoltaic applications.

Joshua M. Pearce received his Ph.D. in Materials Engineering from the Pennsylvania State University. He then developed the first Sustainability program in the Pennsylvania State System of Higher Education as an assistant professor of Physics at Clarion University of Pennsylvania and helped develop the Applied Sustainability graduate program while at Queen's University, Canada. He currently is an Associate Professor cross-appointed in the Department of Materials Science & Engineering and in the Department of Electrical & Computer Engineering at the Michigan Technological University where he runs the Michigan Tech Open Sustainability Technology (MOST) Lab. His research spans electronic device physics, materials engineering of solar photovoltaic cells, and applied sustainability.

## Figure and Table Captions

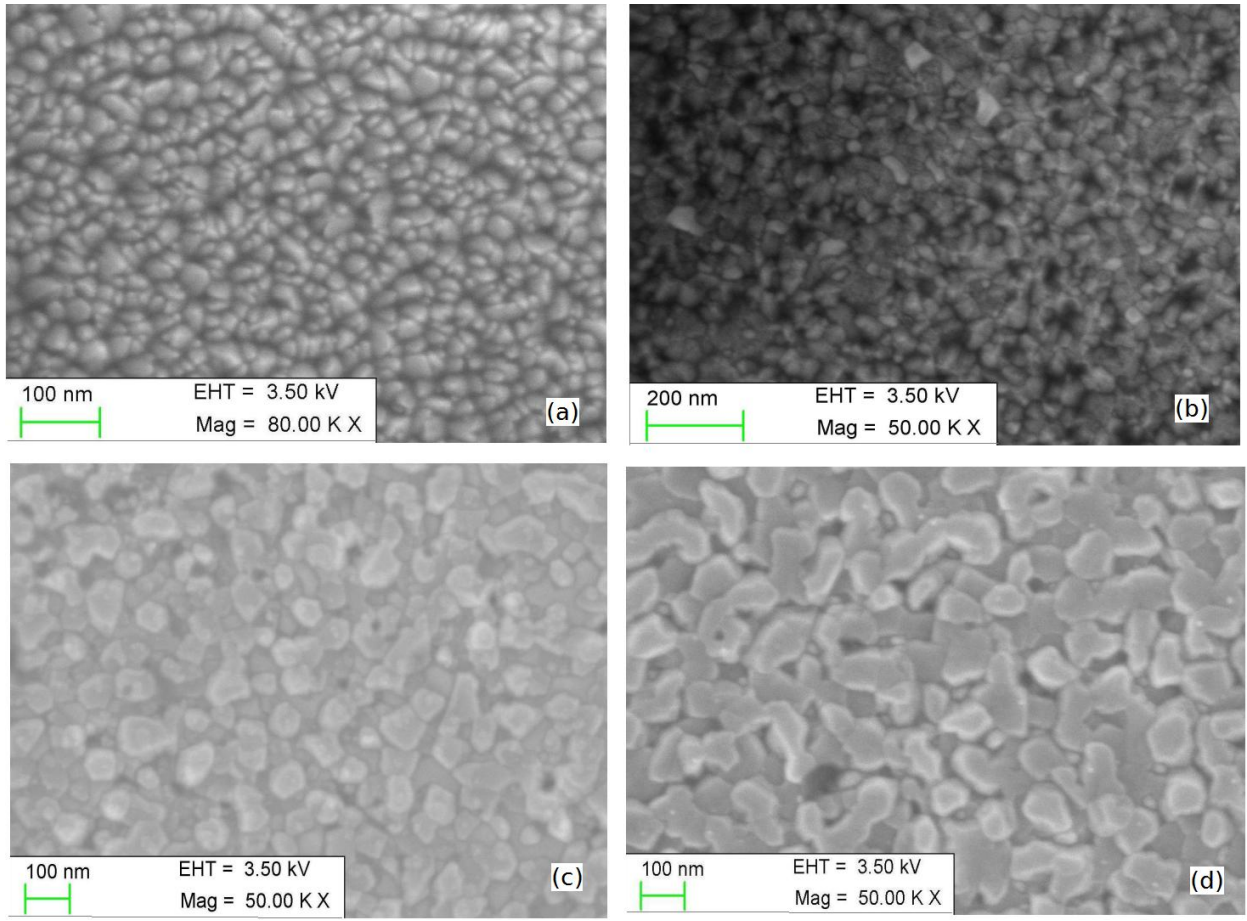

**Fig 1:** Top-down SEM images for a (a) 177 nm-thick $In_{0.38}Ga_{0.62}N$ film; a (b) 202 nm-thick $In_{0.54}Ga_{0.46}N$ film; a (c) 204 nm-thick $In_{0.64}Ga_{0.36}N$ film; and a (d) 221 nm-thick $In_{0.68}Ga_{0.32}N$ film on silicon dioxide substrates with thin GaN buffer layers.

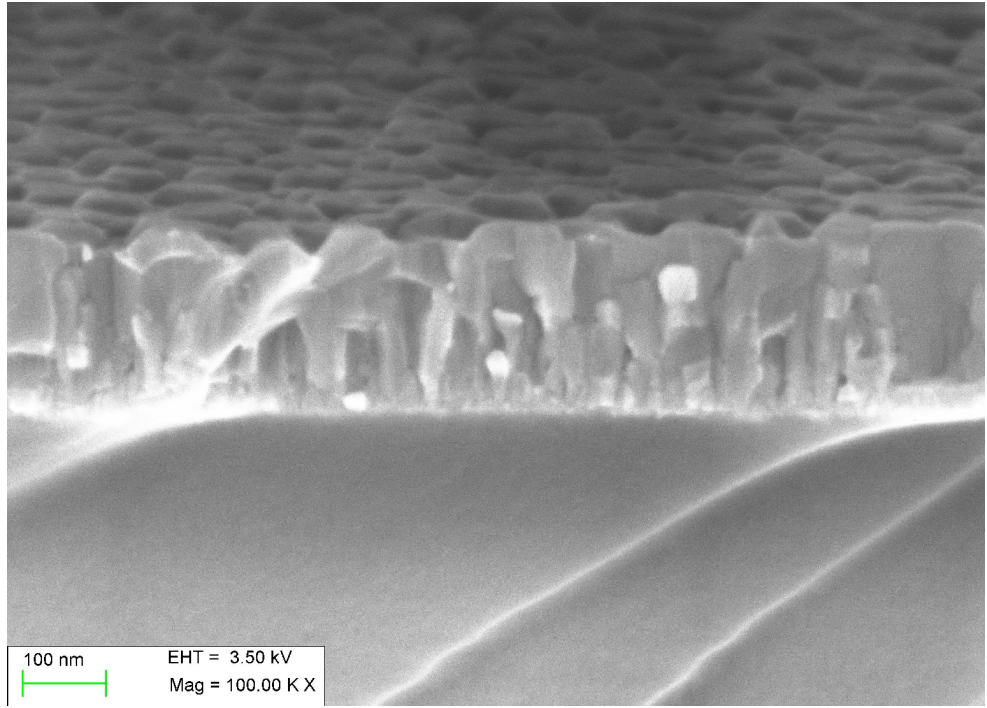

**Fig 2:** Cross-sectional SEM image for the 221 nm-thick $In_{0.68}Ga_{0.32}N$ film on a silicon dioxide substrate with a thin GaN buffer layer.

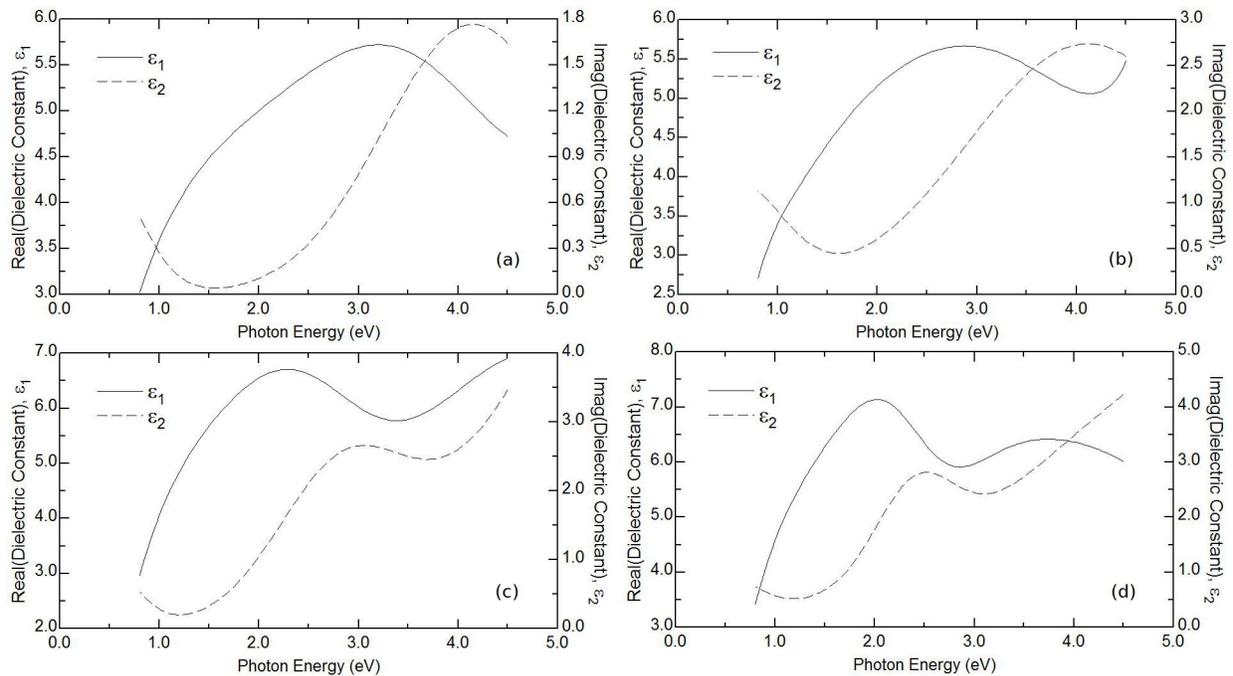

**Fig 3:** Real and imaginary parts of the dielectric function for a (a) 177 nm-thick $In_{0.38}Ga_{0.62}N$ film; a (b) 202 nm-thick $In_{0.54}Ga_{0.46}N$ film; a (c) 204 nm-thick $In_{0.64}Ga_{0.36}N$ film; and a (d) 221 nm-thick $In_{0.68}Ga_{0.32}N$ film on silicon dioxide substrates with thin GaN buffer layers.

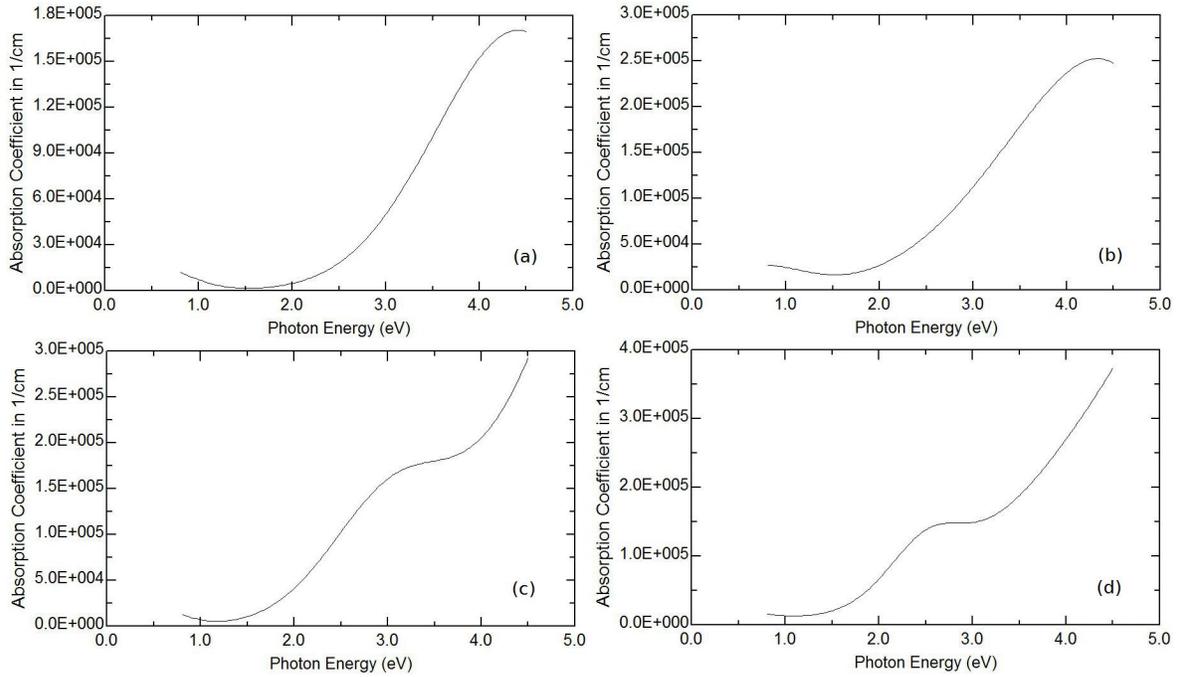

**Fig 4:** Absorption coefficients for a (a) 177 nm-thick $In_{0.38}Ga_{0.62}N$ film; a (b) 202 nm-thick $In_{0.54}Ga_{0.46}N$ film; a (c) 204 nm-thick $In_{0.64}Ga_{0.36}N$ film; and a (d) 221 nm-thick $In_{0.68}Ga_{0.32}N$ film on silicon dioxide substrates with thin GaN buffer layers.

| Composition | $In_xGa_{1-x}N$ Layer Thickness (nm) | Bruggeman Layer Thickness (nm) | Total Film Thickness (nm) |
|---|---|---|---|
| $In_{0.38}Ga_{0.62}N$ | 157 | 20 | 177 |
| $In_{0.54}Ga_{0.46}N$ | 191 | 11 | 202 |
| $In_{0.64}Ga_{0.36}N$ | 189 | 15 | 204 |
| $In_{0.68}Ga_{0.32}N$ | 205 | 16 | 221 |

**Table 1:** $In_xGa_{1-x}N$ and Bruggeman layer thicknesses determined from spectroscopic ellipsometry modelling.

**Table 2:** Parameters for the Gaussian oscillators, amplitude (A), centering energy ($E_n$) and standard deviation ($\sigma$), used to model each $In_xGa_{1-x}N$ film.

| | $In_{0.38}Ga_{0.62}N$ | $In_{0.54}Ga_{0.46}N$ | $In_{0.64}Ga_{0.36}N$ | $In_{0.68}Ga_{0.32}N$ |
|---|---|---|---|---|
| $A_1$ | 1.77 | 2.74 | 9.78 | 5.47 |
| $En_1$ | 4.15 | 4.11 | 6.33 | 6.03 |
| $\sigma_1$ | 0.9 | 1.2 | 1.43 | 0.13 |
| $A_2$ | 2.31 | 2.09 | 1.99 | 1.47 |
| $En_2$ | 0.14 | 0.29 | 2.8 | 2.4 |
| $\sigma_2$ | 0.45 | 0.57 | 0.67 | 0.39 |
| $A_3$ | - | - | 1.69 | 1.26 |
| $En_3$ | - | - | 0.23 | 0.26 |
| $\sigma_3$ | - | - | 0.37 | 0.42 |